\documentclass[a4paper]{JACoW-GSI}
\usepackage{graphicx}
\usepackage{url}
\usepackage{amsmath}
\usepackage{amssymb}
\usepackage{amsfonts}
\usepackage{braket}

\begin{document}
\title{Theory of Semileptonic Charm Decays}

\author[1,2]{J. F. Kamenik\thanks{jernej.kamenik@lnf.infn.it}}
\affil[1]{INFN, Laboratori Nazionali di Frascati, I-00044 Frascati, Italy}
\affil[2]{J.\ Stefan Institute, 1000 Ljubljana, Slovenia}

\maketitle


\section{Introduction}

Over the recent years, the studies of semileptonic charm decays have steadily been (re)gaining importance. Within the standard model (SM)  the CKM matrix element moduli $|V_{cs}|$ and $|V_{cd}|$ are currently best constrained indirectly, through CKM unitarity, and not by direct measurement.  Recently however, deviations from unitarity predictions have been reported in leptonic decays of the $D_s$ meson. Since these processes are helicity-supressed in the SM, the complementary semileptonic channels might offer the opportunity to enlighten the issue. Like in the kaon and $B$-meson decays it is essential to reduce
the theoretical error on the leptonic and semileptonic decays, in order to
match the current and future experimental accuracy and therefore look for the potential discrepancies between the direct and indirect  $|V_{cq}|$ determinations.

In addition, semileptonic charm decays offer a useful testing ground for theoretical tools. 
The bulk of the measured experimental charm events is represented by non-leptonic
$D$-decays, a consistent QCD-based description of which is still missing.
Only a small fraction consists of leptonic and semileptonic decays which
we know how to describe and --at least in principle-- compute from
the first principles of QCD. In this endeavor, numerical lattice QCD calculations need additional theoretical inputs to control systematic uncertainties and to be able to connect to the experimental measurements. Conversely, calculations based on heavy quark expansion (HQE) and operator product expansion (OPE) need to control power corrections, which are of utmost importance for example in  the extraction of $|V_{ub}|$ from inclusive $B$-meson decays, and are expected to be far more pronounced in inclusive charm decays.

\section{Exclusive semileptonic decays}

\subsection{Motivation}
Within the SM the $V_{cs}$ and $V_{cd}$ CKM moduli 
can be determined from CKM unitarity~\cite{CKMFitter09}
\begin{subequations}
\begin{eqnarray}
|V_{cd}|_{UT}  &=&  0.22508 \pm 0.00082\,, \\
|V_{cs}|_{UT} & = & 0.97347 \pm 0.00019\,.
\end{eqnarray}
\end{subequations} 
On the other hand, these quantities can also be extracted directly from recent experimental measurements of leptonic decays $D^- \to \mu \bar\nu$~\cite{:2008sq} and $D_s \to \mu(\tau) \bar\nu$~\cite{Alexander:2009ux, Onyisi:2009th, :2007ws}.  Using the averages of Cleo \& Belle measurements taken fom~\cite{Akeroyd:2009tn} and the most precise decay constants calculations on the lattice~\cite{Follana:2007uv, Blossier:2009bx} one obtains
\begin{subequations}
\begin{eqnarray}
|V_{cd}|_L  &=& |V_{cd}|_{UT} (1.00 \pm 0.05)\,, \\
|V_{cs}|_L   &=& |V_{cs}|_{UT} (1.08 \pm 0.03)\,, 
\end{eqnarray}
\end{subequations} 
in particular, there is a $2.3\sigma$ tension between such determination of $|V_{cs}|$ and the CKM unitarity fit. This calls for cross-checks of lattice QCD calculations as well as experimental measurements.

\subsection{$ D \to P \ell \bar\nu$}

The differential decay width of the semileptonic decays of charmed ($D, D_s$) mesons to light pseudoscalar mesons ($P=\pi, K, \eta^{(')}$) and light leptons ($\ell=\mu$, $e$) in the SM can be parametrized in terms of two kinematical variables
\begin{equation}
\frac{d\Gamma(D\to P \ell \bar \nu)}{dq^2 d\cos \theta_\ell} = \frac{G_F^2 |V_{cq}|^2}{32\pi^3} {|\bf p|^3} |f_+(q^2)|^2 \sin^2 \theta_\ell\,,
\label{eq:SLG}
\end{equation}
where $q^2=(p'-p)^2$ is the momentum exchanged squared, $|{\bf p}| = \sqrt{\lambda(m_D^2,q^2,m_P^2)}/ 2 m_D$ is the absolute three-momentum of the final-state meson in the $D$ rest-frame [$\lambda(a,b,c)=a^2+b^2+c^2-2(ab+bc+ac)$] and $\theta_\ell$ is the angle between the directions of final state meson and lepton again in the $D$ rest-frame. 
\begin{figure}[t]
\begin{center}
\includegraphics[width=8.4cm]{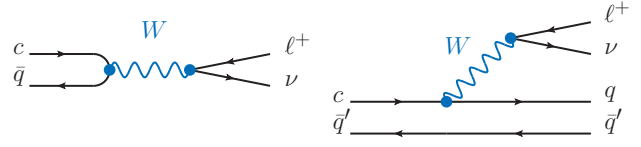}
\end{center}
\caption{The tree-level charged-current process contributing to leptonic (left) and exclusive semileptonic (right) $D$ decays}%
\label{fig:1}%
\end{figure}
The non-perturbative QCD dynamics is encoded in the relevant quark current matrix element and can be parametrized in terms of the suitable form factors, e.g.
\begin{eqnarray}
\bra{P(p)} \bar q \gamma^\mu c \ket{D(p')} \hspace{-0.1cm} &=&  \hspace{-0.1cm} f_+(q^2) \hspace{-0.cm}\left[P^{\mu} - \frac{m_D^2-m_P^2}{q^2} q^\mu\right] \nonumber \\
&& \hspace{-0.1cm} +\,  f_0(q^2) \frac{m_D^2-m_P^2}{q^2} q^\mu\,,
\end{eqnarray}
where $P\equiv p'+p$. In the SM the contribution of $f_0$ to the decay rate is (helicity) suppressed by the charged lepton mass and has thus been neglected in eq. (\ref{eq:SLG}). The relevant CKM matrix element $|V_{cq}|$ can be extracted from experimental measurement of the (partial) decay width provided (1) the  
form factor normalization is known at a single kinematical point 
(customarily at $q^2 = 0$); (2)
a (partial) phase space integral can be extracted and compared to experiment~\cite{:2008yi}.
The form factor normalization can be calculated e.g. using QCD (light-cone) sum rules~\cite{Ball:2006yd}, 
the precision of which however is intrinsically limited. Therefore a more accurate determination can only possibly be expected from lattice QCD simulations.  

In the past, lattice calculations were only accurate in a kinematical 
region of diminishing phase-space near $q^2_{max} \equiv (m_D-m_P)^2$. 
A subsequent extrapolation between lattice and experiment was needed using suitable form factor parameterizations. Recent lattice studies however can already cover the complete kinematical region in D decays~\cite{AlHaydari:2009zr, Aubin:2004ej, Abada:2000ty}. Extrapolation procedures nevertheless remain important in exclusive determinations of $|V_{ub}|$ from $B\to\pi$ transitions with considerably larger phase space. It is illustrative to consider the present experimental precision of the measured 
$B \to \pi \ell \bar \nu $ partial branching fractions with a cut on $q^2$ close to where lattice QCD studies can provide the form factor normalization~\cite{HFAG08}
\begin{subequations}
\begin{eqnarray}
BF(total) \hspace{-0.2cm}&=&\hspace{-0.2cm} 1.36 \times 10^{-4} (1 \pm 0.05)\,, \\ 
BF(q^2<16 \mathrm{GeV}^2) \hspace{-0.2cm}&=&\hspace{-0.2cm} 0.94 \times 10^{-4} (1 \pm 0.07)\,, \\ 
BF(q^2>16 \mathrm{GeV}^2) \hspace{-0.2cm}&=&\hspace{-0.2cm} 0.37 \times 10^{-4} (1 \pm 0.10)\,.
\end{eqnarray}
\end{subequations}
By only considering the high $q^2$ region, the experimental precision is reduced by as much as $40\%$. In order to take advantage of the complete available experimental statistics, a careful control of theoretical errors, associated with form factor parameterizations is needed. These in term can be tested in the charm sector.

Exact (analytic) shape of the form factors is illusive, as it is non-perturbative in nature. Lattice QCD can be used to calculate the values of the form factors numerically at individual $q^2$ points, leaving the question of how to best (extra-)interpolate between the various $q^2$ regions open. Most form factor parameterizations employ an expansion in functions of $q^2$.  
\begin{itemize}
\item The simplest Taylor expansion around $q^2=0$ has proven useful in $K\to\pi$  transitions, where the required convergence radius is small $q_{max}^2 = (m_K-m_\pi)^2$.   It is however less suitable for semileptonic decays of heavy mesons, where the phase space is much larger. Perhaps even more important however is the presence of nearby resonance poles, close to the physical region in $D\to P$ (and $B\to P$) transitions.
\item An orthogonal approach is represented by the sum over t-channel resonance pole contributions of the form $(q^2-m_i^2)^{-1}$. It has been traditionally used both in theoretical form factor calculations as well as in the experimental fits.
\item Using a conformal mapping of the complex $t$ plane and taking into account the position of the physical cut at $ t_+ = (m_D+m_P)^2$, one can generalize the expansion around any point $t_0$ in terms of $z(q^2,t_0) =  (\sqrt{t_+-q^2} - \sqrt{t_+-t_0})/(\sqrt{t_+-q^2}+\sqrt{t_+-t_0})$. The required convergence radius is bounded by $ [0,|z_{max}|<1]$  (c.f.~\cite{Boyd:1997qw}).
\end{itemize}
Some additional information is provided by the form factor dispersion relations.  Namely, in $D \to K$ decays the first t-channel resonance ($D_s^*$) is below the 
physical cut at $q^2 = t_+$  (the same reasoning holds also for $B\to\pi$ transitions with the $B^*$ resonance)
\begin{subequations}
\begin{eqnarray}
f_0(q^2) \hspace{-0.2cm} &=& \hspace{-0.2cm} \frac{1}{\pi} \int_{t_+}^\infty d t \frac{\mathrm{Im}[ f_0(t)]}{t-q^2-i\epsilon}\,,\\
f_+(q^2) \hspace{-0.2cm}&=&\hspace{-0.2cm} \frac{\mathrm{Res}[ f_+(q^2)]_{q^2=m_{D_s^*}^2}}{m_{D_s^*}^2-q^2} + \frac{1}{\pi} \int_{t_+}^\infty d t \frac{\mathrm{Im}[ f_+(t)]}{t-q^2-i\epsilon}\,.\nonumber\\
\end{eqnarray}
\end{subequations}
Close to the kinematical end-point at $q^2_{max}$, the $f_+$ form factor is therefore expected to be completely dominated by the first resonance pole term (see also~\cite{DescotesGenon:2008hh}).

Finally,  in the limit $m_c \to \infty$, $m_q \to 0$, the form factors obey the so-called heavy quark scaling relations. When the final state meson is soft in the $D$ rest-frame (near $q^2_{max}$), heavy quark effective theory predicts~\cite{Isgur:1990kf}
\begin{subequations}
\begin{eqnarray}
f_+(q^2 \approx q^2_{ max},m_D) &\sim& m_D^{1/2}\,,\\
f_0\, (q^2 \approx q^2_{max},m_D) &\sim& m_D^{-1/2}\,.
\end{eqnarray}
\end{subequations}
On the other end of the kinematical phase space (near $q^2=0$) similar scalling relations have been initially found by using QCD sum rules~\cite{Chernyak} 
\begin{eqnarray}
f_{+,0}(q^2 \approx0,m_D) \sim m^{- 3/2}_D\,.
\end{eqnarray}
They were subsequently confirmed within the soft collinear effective theory (SCET)~\cite{Charles:1998dr, Beneke:2000wa} which in this limit predicts also
\begin{eqnarray}
f_0 \approx \frac{2 E_P}{ m_D} f_+ \,,
\end{eqnarray}
where $E_P$ is the energy of the outgoing meson in the rest-frame of the initial $D$ meson. These relations are broken by (potentially large) perturbative $\alpha_s(m_c)$, and power ($\Lambda/m_c$) corrections.

Phenomenologically, the single pole ansatz for $f^+(q^2)$
\begin{equation}
f_+ (q^2) = \frac{f^+ (0)}{ (1 -x)}\, ,\quad x=q^2/m^2_{pole}\,,
\end{equation}
has been found inconsistent with measured spectra of $D\to \pi \ell \nu$ and $D\to K \ell\nu$ already some time ago~\cite{Link:2004dh, Huang:2004fra}.

A simple three-parameter ansatz respecting leading-order HQ scaling relations was constructed by Be\'cirevi\'c \& Kaidalov (BK)~\cite{Becirevic:1999kt}  
\begin{equation}
f_+(q^2)= \frac{f(0)}{ (1 -x)(1-ax) }, \quad f_0(q^2)=\frac{ f(0)}{ (1 -bx)}\,.
\end{equation}
Considered as a truncated pole expansion, by construction it assumes 
\begin{equation}
\frac{1}{f_+(0)}\left[ \frac{d f_+(x)}{dx} - \frac{d f_0(x)}{dx} \right]\Big|_{x\to 0} \approx 1
\end{equation}
or  equivalently $a\approx b\, [\sim \mathcal O(1)]$. These assumptions were found not to hold for $D\to\pi$, $D\to K$ decays~\cite{Hill:2006ub, Fajfer:2004mv}. On the other hand, at present precision the data are still well represented by an unconstrained BK fit~\cite{:2008yi}.   

Using dispersion relations and conformal $t$-plane mapping, an alternative description of the form factor shape can be constructed~\cite{Boyd:1997qw}
\begin{equation}
f_+(q^2) = \frac{1}{P(q^2)\phi(q^2,t_0)} \sum_{k=0}^\infty a_k(t_0) [z(q^2,t_0)]^k\,,
\end{equation}
where $P(q^2)$ subtracts away poles below the continuum threshold at $t_+$ and
$\phi(q^2,t_0)$ is a normalization factor conventionally fixed from perturbative OPE unitarity (see also~\cite{Bourrely:2008za}). Such parameterizations have been used extensively to describe semileptonic $B$ decays. In $D\to P$ transitions, existing data has already proven much more constraining than unitarity bounds~\cite{Becher:2005bg}. Nonetheless, by comparing to experiment~\cite{:2008yi} while using the form factor normalization from the lattice~\cite{Aubin:2004ej} one obtains for the relevant CKM moduli~\cite{CKMFitter09}
\begin{subequations}
\begin{eqnarray}
|V_{cd}|_{SL} &=&  |V_{cd}|_{UT} (0.99 \pm 0.11)\,, \\
|V_{cs}|_{SL} & =& |V_{cs}|_{UT} (1.05 \pm 0.13)\,. 
\end{eqnarray}
\end{subequations}
At present precision (dominated by the theoretical errors on the lattice form factor normalization) the values obtained in this way are well consistent with CKM unitarity, although the central value of $|V_{cs}|$ is larger than one, as in the case of leptonic decays. Finally, once the theoretical lattice QCD error will approach the present experimental sensitivity, the consistency of the applied form factor parameterizations will also be put to the test.

Beyond the SM, new operator contributions may contribute signiÞcantly to the decay rate 
 \begin{subequations}
\begin{eqnarray}
\bra{P(p)} \bar q \sigma^{\mu\nu} c  \ket{D(p')} \hspace{-0.2cm}&=& \hspace{-0.2cm} \frac{ i  ({P}^{\mu} q^{\nu} - {P}^{\nu} q^{\mu})}{(m_D+m_P)} f_T(q^2)\,, \nonumber\\
&&\\
\bra{P(p)} \bar q c  \ket{D(p')} \hspace{-0.2cm}&=&\hspace{-0.2cm} \frac{m_D^2-m_P^2}{m_c-m_q} f_0(q^2)\,.
\end{eqnarray}
\end{subequations}
To describe possible tensor contributions, the knowledge of one additional form factor is needed ($f_T$). In both SCET \& HQET limits~\cite{Isgur:1990kf, Beneke:2000wa} however one finds~\cite{Isgur:1990jg, Hill:2005ju, Bartsch:2009qp}
\begin{equation}
f_T \approx  f_+\,.
\label{eq:FT}
\end{equation}
The relation can be traced back to heavy quark spin symmetry relating the matrix elements of the $\sigma_{\mu\nu}$ and $\gamma_\mu$ Dirac structures acting on a heavy quark field. However it also requires the HQ and SCET form factor scaling laws to hold. In addition, the $f_T$ form factor depends on the QCD anomalous dimension of
the tensor operator (whereas the conserved vector current does not). Therefore one can only speculate that the relation should approximately hold throughout the kinematical region at some operator matching scale close to the charm mass. The proposal can of course be tested on the lattice.

Phenomenologically, new operators may contribute differently to leptonic and semileptonic decays. 
In particular, while chiral current interactions produce similar enhancement in both leptonic and semileptonic modes (proportional to the $f_+$ form factor in the semileptonic case), scalar or tensor interactions need to scale with the lepton mass in order to produce similar enhancement in $\tau$ and $\mu$ leptonic modes as indicated by experiments. This means that such interactions will contribute negligibly to the semileptonic decay rates ($\tau$ 
channel is kinematically forbidden).  Other observables, available in the semileptonic mode may help to discriminate among contributions. A particular asymmetry was constructed in~\cite{Kronfeld:2008gu}
\begin{equation}
\mathcal A_\bot = \frac{\Gamma (E_{\ell\bot}>0) - \Gamma (E_{\ell\bot}<0)  }{\Gamma (E_{\ell\bot}>0) +\Gamma (E_{\ell\bot}<0) }\,,
\label{eq:AS}
\end{equation}
where 
\begin{equation}
E_{\ell\bot} = E_\ell -\frac{1}{2} (m_D-E_P) (1+m_\ell^2/q^2)\,,
\end{equation}
and $E_\ell$ is the energy of the outgoing charged lepton in the rest-frame of the initial $D$ meson. This observable is directly sensitive to the interference between SM and scalar or tensor interactions.  A dedicated experimental analysis would possibly need to determine the feasibility of such a measurement. 

\subsection{$D\to V \ell \bar \nu$}

Traditionally the semileptonic decays of charmed mesons to light vector mesons ($V=\rho,K^*,\omega,\phi$) were considered less interesting due to the larger number of the form factors needed to describe the decay rate, less theoretical control over their evaluation, as well as more challenging experimental analyses. On the other hand, the various polarizations of the vector meson in the final state give access to more observables suitable for distinguishing between SM and possible new physics contributions. 

In the SM (as well as in presence of new scalar contributions) the $D\to V$ transitions can be fully described in terms of four form factors $V$, $A_{0,1,2}$ defined through
\begin{eqnarray}
	\langle V (\epsilon,p) | \bar q \gamma^{\mu} c | D (p') \rangle \hspace{-0.1cm}&=&\hspace{-0.1cm} \frac{2 V(q^2)}{m_D + m_V} \epsilon^{\mu\nu\alpha\beta} \epsilon_{\nu}^* p'_{\alpha} p_{\beta}, \nonumber\\*
	\langle V (\epsilon,p) | \bar q \gamma^{\mu} \gamma^5 c | D (p') \rangle \hspace{-0.1cm}&=&\hspace{-0.1cm}  - i \epsilon^* \cdot q \frac{2m_V}{q^2} q^{\mu} A_0(q^2) \nonumber\\
	&&\hspace{-2.3cm}- i(m_D + m_V) \left[\epsilon^{*\mu} - \frac{\epsilon^{*} \cdot q}{q^2} q^{\mu}\right] A_1(q^2) \nonumber\\*
	&&\hspace{-3.2cm}+ i \frac{\epsilon^{*}\cdot q}{(m_D + m_V)}\left[P^{\mu} - \frac{m_D^2-m_V^2}{q^2} q^{\mu}\right] A_2(q^2),\nonumber\\* 
\end{eqnarray}
If additional tensor contributions are present, these require the knowledge of additional three form factors $T_{1,2,3}$ 
\begin{eqnarray}
\langle V(p,\epsilon )\vert \overline q\sigma^{\mu\nu}c \vert D(p')\rangle
&=& -i \epsilon^{\ast}_\alpha  {\cal T}^{\alpha\mu\nu }\,,\cr 
&&\cr
 &&\hspace*{-4cm}{\cal T}^{\alpha\mu\nu }= \epsilon^{ \alpha\mu \nu \beta} 
\left[
\left(  P_\beta - {m_D^2 - m_V^2 \over q^2 } q_\beta \right) T_1(q^2)\right.\cr
&&\hspace{0.1cm}\left.+ {m_D^2 - m_V^2 \over q^2 } q_\beta 
 T_2(q^2) \right]  \cr
&&\hspace*{-2.8cm} + {2 {p'}^\alpha \over q^2}  \epsilon^{\mu \nu \sigma \lambda} 
p^{\prime}_\sigma p_{\lambda} \bigg[ T_2(q^2)- T_1(q^2) \cr
&&\hspace{0.4cm}\left. + {q^2\over m_D^2-m_V^2}T_3(q^2) \right] .\cr
 &&
\end{eqnarray}
While lattice QCD computations can provide normalization of the form factors at various kinematical points, HQ scaling laws and relations can again be used together with dispersion relations to construct useful parameterizations. In the soft $V$ limit~\cite{Isgur:1990kf}
\begin{subequations}
\begin{eqnarray}
V(q^2 \approx q^2_{max},m_D) &\sim& m_D^{1/2}\,,\\
A_{0,2}(q^2 \approx q^2_{max},m_D) &\sim& m_D^{1/2}\,,\\
A_1(q^2 \approx q^2_{max},m_D) &\sim& m_D^{-1/2}\,,\\
T_{1,3}(q^2 \approx q^2_{max},m_D) &\sim& m_D^{1/2}\,,\\
T_2(q^2 \approx q^2_{max},m_D) &\sim& m_D^{-1/2}\,,
\end{eqnarray}
\end{subequations}
while near $q^2\approx 0$  all form factors should scale as $m_D^{-3/2}$~\cite{Charles:1998dr, Beneke:2000wa}.
There are also additional form factor relations, broken by $\alpha_s(m_c)$, $\Lambda/m_c$ corrections, which allow analogues of the BK parameterization for $H\to V$ decays to be constructed~\cite{Fajfer:2005ug, Becirevic:2006nm}
\begin{subequations}
\begin{eqnarray}
\label{eq:V}
V(q^2)&=& \frac{V(0)}{ (1 -x)(1-ax)} \,,\\
A_0(q^2)&=& \frac{A_0(0)}{ (1 -y)(1-a'y)}\,,\\ 
A_1(q^2)&=& \frac{A_1(0)}{ (1 -bx)} \,,\\
A_2(q^2)&=& \frac{A_2(0)}{ (1 -bx)(1-b'x)}\,,\\
\label{eq:T}
T_1(q^2)&=& \frac{T(0)}{ (1 -x)(1-ax)} \,,\\
T_2(q^2)&=& \frac{T(0) }{(1 -bx)}\,,\\
T_3(q^2)&=& \frac{T_3(0)}{ (1 -bx)(1-b'x)}\,,
\end{eqnarray}
\end{subequations}
with, $V(0)/A_1(0)/T(0) \approx 1, a^{(i)} \approx b^{(i)} \sim \mathcal O(1) $. The different nearest resonance poles in $V,\ A_i$ and $T_i$ are reflected in the different normalization of the variables $x,y=q^2/m_{pole}^2$. Note that eqs. (\ref{eq:V}) and (\ref{eq:T}) already tacitly hint at the relation
\begin{equation}
T_1 \approx V\,,
\end{equation}
throughout the allowed kinematic region, which can be traced back to the same origins as eq. (\ref{eq:FT}) but is expected to receive additional large corrections due to non-negligible light vector masses.
The differential decay width is usually written in terms of Helicity amplitudes $H_{+-0}$ 
\begin{eqnarray}
\frac{d \Gamma(D\to V \ell \bar \nu)}{dq^2 d\cos \theta_\ell} \hspace{-0.3cm}&=& \hspace{-0.3cm}\frac{G_F^2|V_{cq}|^2}{128\pi^3m_D^2} |{\bf p}^*| q^2  \left[ \frac{(1-\cos\theta_\ell)^2}{2} |H_-|^2\right.\nonumber\\
&&\hspace{-1.1cm} \left. + \frac{(1+\cos\theta_\ell)^2}{2} |H_+|^2 + \sin^2\theta_\ell |H_0|^2  \right]\,,
\end{eqnarray}
which can in term be related to certain combinations of the various $D\to V$ form factors.
Experimental information on the shapes of $H_{+-0}$ is already available~\cite{Link:2005dp, Shepherd:2006tw, Wiss:2007mr} and can be used to test form factor parameterizations like the z-expansion~\cite{Hill:2006ub} or the modified (BK) pole ansatz~\cite{Fajfer:2006uy}.
In addition to the (partially) integrated decay rate, suitably constructed asymmetries, analogous to (\ref{eq:AS}) might provide more direct access to anomalous scalar, tensor contributions.

\section{Inclusive Semileptonic 
Decays}

\subsection{Introduction}

Over the last years, there has been a tremendous progress in the determination of the $|V_{ub}|$ CKM matrix element from the measurement of the inclusive decay rate $B\to X_u \ell\bar\nu$. Using OPE and HQE the theoretical predictions~\cite{Lange:2005yw, Andersen:2005mj, Gambino:2007rp, Aglietti:2007ik, Bauer:2001rc} have reached a precision below 10\%~\cite{HFAG08}. Estimation of relevant power-supressed non-perturbative operator matrix element values from inclusive analyses of $B\to X_c\ell\bar\nu$ has been instrumental in this effort. At the third order in the heavy quark mass expansion ($1/m_b^3$) however, effects due to dimension-6 four quark 
operators, the so-called weak annihilation (WA) contributions appear,
which cannot be extracted from 
the inclusive $b\to c$ analysis. 
In addition, these are $16\pi^2$ phase space enhanced 
compared to LO \& NLO 
contributions (such enhancement does does not seem to appear at dimension-7 
~\cite{Dassinger:2006md}).

Recently, the inclusive semileptonic decay rates of charmed mesons ($D \to X \ell \bar\nu$ 
) have been determined experimentally, yielding for the branching fractions~\cite{Adam:2006nu, Ablikim:2008qp}
 \begin{subequations}
\begin{eqnarray}
B(D^+ 
\to Xe\bar\nu) \hspace{-0.2cm}&=&\hspace{-0.2cm} (16.13\pm 0.20\pm 0.33)\%\,, \\
B(D^0 
\to Xe\bar\nu) \hspace{-0.2cm}&=&\hspace{-0.2cm} (6.46\pm 0.17\pm 0.13)\%\,,
\end{eqnarray}
\end{subequations}
with similar results for muons. 
On the theory side, one may attempt to accommodate these numbers by treating the charm quark mass as heavy, and perform an OPE calculation by expanding in $\alpha_s(m_c)$, $\Lambda/m_c$.  In the process, one needs to estimate contributions of local operator matrix elements, which can possibly be related to the ones appearing in inclusive $B$ decay analyses or estimated on the lattice.

\subsection{Heavy quark and operator product
expansion}

\begin{figure}[!tch]
\begin{center}
\includegraphics[width=8.2cm]{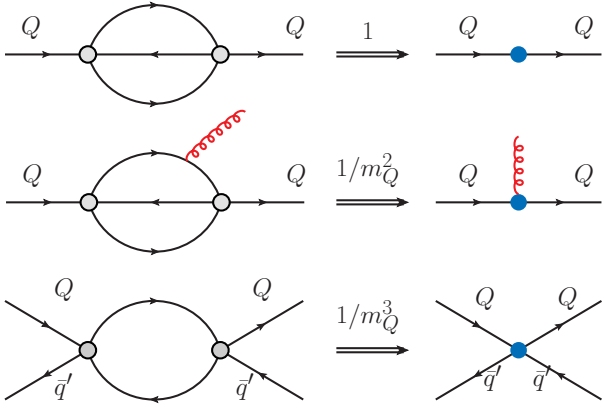}
\end{center}
\caption{Examples of contributions to the transition operator 
$\mathcal T$ (left) and to the 
corresponding local operator (right). The open circles represent the insertions of the 
weak effective Hamiltonian. The full circles represent the insertion of a local $\Delta Q = 0$ 
operator.}%
\label{fig:2}%
\end{figure}
The total inclusive decay rate can be related to the absorptive part of the forward scattering amplitude by applying the optical theorem 
\begin{equation}
\Gamma(H_{Q \bar q}) = \frac{1}{2m_H} \bra{H_{Q\bar q}} \mathcal T \ket{H_{Q\bar q}}\,,
\end{equation}
where 
\begin{equation}
\mathcal T = \mathrm{Im}\left\{ i \int d^4 x\, \mathrm{T} [ \mathcal H_{eff}(x) \, \mathcal H_{eff}(0) ] \right\}\,,
\end{equation}
and $\mathrm{T}$ is the time-ordering operator while $\mathcal H_{eff}$ is the effective weak Hamiltonian mediating charged current semileptonic processes. 
For a decaying hadron containing a single heavy quark one can, assuming quark-hadron duality, evaluate the quantity by expanding it in powers of the inverse heavy quark mass, systematically including perturbative QCD corrections. Applying the strategy to semileptonic $D$ meson decays one obtains~\cite{Bigi:1992su, Manohar:1993qn}
\begin{eqnarray}
&&\hspace{-0.5cm}\Gamma(D\to X \ell \bar \nu) = \frac{G_F^2 m_c^5}{192 \pi^3} |V_{cs}|^2 g(r) \nonumber\\
&&\hspace{-0.5cm}\times \left\{ \frac{1}{2m_D}\left[ I_0(r) \bra{D} \bar c c \ket{D} + \frac{I_1(r)}{m_c^2} \bra{D} \bar c g_s \sigma \cdot G c \ket{D} \right.\right.\nonumber\\
&&\hspace{-0.5cm}\left.\left. - \frac{16\pi^2}{2m_c^3}\bra{D} \mathcal O_{V-A} - \mathcal O_{S-P} \ket{D} + \ldots \right.] \right\}\,,
\label{eq:OPESL}
\end{eqnarray}
where $r=m_s^2/m_c^2$, $g_s$ is the QCD coupling constant, $G_{\mu\nu}=[\mathcal D, \mathcal D]$, where $\mathcal D$ is the QCD covariant derivative, $g(r)$ is the leading kinematical factor, while $I_{0,1}(r)$ contain relative kinematical as well as perturbative $\alpha_s(m_c)$ corrections. The ellipses denote additional $1/m_c^3$ and higher power corrections which are not phase-space enhanced.

The non-perturbative operator matrix elements appearing in eq. (\ref{eq:OPESL}) can be parameterized using the heavy quark equations of motion 
\begin{equation}
\bar c  c = \bar c v \hspace{-0.17cm}\backslash c + \frac{1}{2m_c^2} \left[ \bar c (i \mathcal D_\bot)^2 c + \bar c \frac{g_s}{2}\sigma \cdot G c \right] + \mathcal O(1/m_c^3)\,,
\end{equation}
where $\mathcal D_\bot = \mathcal D- v (v\cdot \mathcal D)$, and related to the HQE parameters
\begin{subequations}
\begin{eqnarray}
\mu_\pi^2 &=& -\frac{1}{2m_D} \bra{D} \bar c (i \mathcal D_\bot)^2 c\ket{D} \,,\\
\mu_G^2 &=& \frac{1}{2m_D} \bra{D} \bar c \frac{g_s}{2} \sigma\cdot G c\ket{D} \,,
\end{eqnarray}
\end{subequations}
as
\begin{equation}
\frac{1}{2m_D}\bra{D} \bar c c \ket{D} = 1 - \frac{\mu_\pi^2-\mu_G^2}{2m_c^2}\,.
\end{equation}
Similarly, the operator matrix elements appearing at $1/m_c^3$ can be parameterized in terms of two additional parameters ($\rho_{D,SL}$)~\cite{Bigi:1994ga,Gremm:1996df}. The $\mu_\pi$ and $\rho_D$ parameters depend on the employed heavy quark mass scheme and the associated matching scale~\cite{Bigi:1994ga}.
Finally, also at $1/m_c^3$ the contributions involving light flavors 
\begin{eqnarray}
\mathcal O_{V-A}^{q'} &\equiv&  \bar Q \gamma_\mu (1-\gamma_5) q' \bar q' \gamma^\mu (1-\gamma_5) Q \,, \nonumber\\
\mathcal O_{S-P}^{q'} &\equiv&  \bar Q  (1-\gamma_5) q' \bar q'  (1-\gamma_5) Q  \,.
\label{eq:WA}
\end{eqnarray}
are conventionally parametrized in terms of deviations from the complete factorization or vacuum saturation approximation by introducing the suitable bag parameters
\begin{subequations}
\begin{eqnarray}
\bra{D} \mathcal O_{V-A}\ket{D} &=&f_D^2 m_D^2 B_1 \,, \\
\bra D \mathcal O_{S-P}\ket{D} &=& f_D^2 m_D^2 B_2 \,.
\end{eqnarray}
\end{subequations}
In this way one defines $2m_D B_{\text{WA}}  =  \bra{D}  \mathcal O_{V-A}- \mathcal O_{S-P}\ket{D} = f_D^2 m_D^2 (B_1-B_2)$. These quantities depend on the scale ($\mu_{\text{WA}}$) at which the matrix elements are evaluated and can mix with phase space non-enhanced contributions ($\rho_D$)\cite{Gambino:2005tp}. The early estimates of $B_{1,2}$ were done in the framework of QCD sum rules~\cite{Baek:1998vk, Cheng:1998ia, Gabbiani:2004tp}. They have been also computed on the lattice~\cite{DiPierro:1998ty, Becirevic:2001fy,   Hashimoto:2001zq, Yamada:2001xp}. However, since possible contractions of the $\mathcal O_{V-A,S-P}$ operators involve closed fermionic line topologies (eye-contractions)~\cite{Voloshin:2001xi}, such lattice estimates are necessarily incomplete~\cite{Becirevic:2001fy}.
\begin{figure}[!tch]
\begin{center}
\includegraphics[width=7.cm]{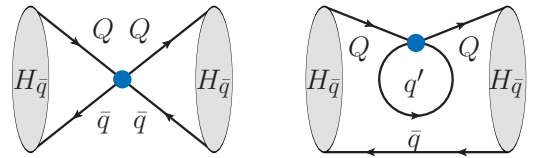}
\end{center}
\caption{Graph topologies of WA operators contributing to inclusive decays of heavy hadrons.  The full circles represent the insertion of a local $\Delta Q = 0$ 
operator. Only the left topology is estimated on the lattice.}%
\label{fig:3}%
\end{figure}

Putting all the pieces together one obtains for the decay rate up to $1/m_c^4$ and $\alpha_s/m_c^2$
\begin{eqnarray}
 &&\hspace{-0.6cm}\Gamma(D\to X \ell \bar \nu) = \frac{G_F^2 m_{c}^5 }{192 \pi ^3} 
   \left|V_{\text{cs}}\right|{   }^2 g(r) \nonumber\\
   &&\times \left\{ 1 +\frac{\alpha _s\,   }{\pi   } p_{\text{1}}(r)+\left(\frac{\alpha _s\,   }{\pi}\right)^2 p_{\text{2}}(r) \right.\nonumber\\
    &&\left. -\frac{\mu _{\pi   }{}^2}{2   m_{c}{}^2}  +\frac{\left(\frac{1}   {2}-\frac{2   (1-r)^4}{g(r)}\right)
   \left(\mu _G^2-\frac{\rho _D{}^3+\rho _{\text{LS}}{}^3}{m_{c}}\right)}{m_{c}   {}^2}  \right.\nonumber\\
   &&\left. + \frac{32 \pi ^2   B_{\text{WA}}(\mu_{\text{WA}})}{m_{c}^3}+\frac{d(r,\mu_{\text {WA}}) \rho   _D{}^3}{g(r)   m_{c}^3}    + \mathcal O (1/m_c^4) \right\}\,,\nonumber\\
   \label{eq:OPESL1}
\end{eqnarray}
where the leading $\mu_{\text{WA}}$ scale dependence (which should cancel between the two terms in the fourth line) is given by  $ d(r,\mu_{\text{WA}}) = [-{10 r^4}+{32   r^3}- 24 r^2- {32   r} + 24 \log   \left({\mu   _{\text{WA}}^2}/{m_c^2}\right)+{34}]/3$. The expressions for $g(r)$ and $p_{1,2}(r)$ can be read from the $X_i$ functions in ref.~\cite{Pak:2008qt} as $g(r)=X_0$, $p_{1,2}(r)=4 X_{1,2}/3X_0$.

\subsection{Phenomenological 
analysis}

The most sensitive parameter entering the inclusive decay rate analysis is the charm quark mass. Without imposing any kinematical cuts it enters the total decay rate with the fifth power. From lattice~\cite{Allison:2008xk}, Charmonium SR~\cite{Kuhn:2007vp, Chetyrkin:2009fv}, $b \to c$ spectral fits~\cite{Buchmuller:2005zv} one obtains a fairly accurate and consistent ($\overline {MS}$) value of
$m_c(m_c)=1.27(2)\mathrm{GeV}$.  
Values of $\mu_{G,\pi}^2$ and $\rho_{D,SL}$ can also be extracted experimentally. Namely $\mu^2_G=(3/4)[m^2_{D^*} -m^2_D]=0.41\mathrm{GeV}^2$, while $\mu^2_\pi 
\approx 0.4\mathrm{GeV}^2$, $\rho_D^3\approx 0.2 \mathrm{GeV}^3$ and $\rho_{SL}^3\approx-0.2 \mathrm{GeV}^3$ are obtained at the $10\% - 20\%$ precision from a fit to the $b\to c$ spectrum~\cite{Hauke:2007mm} in the so-called kinetic heavy mass scheme~\cite{Bigi:1996si} at the scale of 1 GeV.
Adding up all these known contributions in eq. (\ref{eq:OPESL1}), the 
experimental values are saturated to $\approx 70\%$. The calculation exhibits a very slow perturbative \& power convergence, the details will be presented elsewhere~\cite{Gambino}. 
In light of sizable residual scale and scheme dependencies (the best perturbative convergence is achieved for a very low kinetic charm mass scale of 0.5 GeV) the result should thus be taken as tentative at best.  Assuming the unknown enhanced $1/m_c^3$  terms saturate the rate, one can in any case obtain a guestimate~\cite{Becirevic:2008us} of (using $f_D=208(4)\mathrm{MeV}$~\cite{Follana:2007uv}) 
\begin{equation}
(B_1-B_2)(2\mathrm{GeV}) \approx 0.05\,,
\end{equation}
which appears to be a reasonably small number. One should note however, that a fit to the inclusive $D^{+,0}$ decays can only probe spectator see-quark operator contributions (where the flavour $q'$ appearing in  eq. (\ref{eq:WA}) does not match the (light) flavour of the initial charmed hadron), while the valence quark contribution is only relevant in $D_s$ decays. Once determined, this could be related to the $B^+\to X_u \ell^+ \nu$ and $B^0\to X_u \ell^+ \nu$ width difference via $SU(3)$ symmetry~\cite{Voloshin:2001xi}. For more accuracy and cross-checks, improved lattice QCD estimates of $B_{1,2}$ are of course called for.

In the meantime the phenomenological analysis could certainly also be improved by considering higher spectral moments in addition to the total decay rates (as done in $b\to c$ case) in order to leverage more control over the OPE convergence.
Cleo already published the lepton momentum spectra in inclusive semileptonic $D$ decays with a lower cut~\cite{Adam:2006nu} which however requires a more involved treatment. 
Utilizing the moments, a more direct access to power corrections and even possible duality 
violations (c.f.~\cite{Shifman:1994yf, Chibisov:1996wf}) could be obtained, since e.g. WA contributions are expected to dominate near the spectrum end-point, c.f.~\cite{Shifman:1995mt, Benson:2003kp}.

\section{Conclusions}

The exclusive $D \to P \ell \bar\nu$ decays offer the opportunity to confront the recent 
puzzles concerning the determination of $|V_{cq}|$ CKM matrix elements from leptonic charm decays. Here the most critical input from theory is to provide an accurate normalization of the relevant form factors. While new results from lattice QCD calculations are eagerly awaited,  presently the extraction of the afore mentioned CKM elements 
is not yet competitive with leptonic decays.  Furthermore, various form factor parameterizations used to model semileptonic $D$ decay spectra are also employed in $B$ decays. By comparing the measured $D$ decay spectra with lattice results and fitting both to these parameterizations one can test extrapolation 
procedures needed in exclusive determination of $|V_{ub}|$. 
Finally, by using even more of the present experimental information available, anomalous contributions due to scalar, right-handed or  tensor currents could be probed.
Importantly, such an analysis introduces almost no new hadronic uncertainties.
In this respect  the $D \to V \ell \bar \nu$ decays offer potentially even more useful observables, sensitive to NP contributions, with the caveat of heaving to estimate more hadronic form factors.

Inclusive $D \to X \ell \bar\nu$ decays offer the possibility to test and extract power suppressed spectator contributions to inclusive semileptonic decay widths of heavy hadrons from experiment. These are relevant for a reliable extraction of $|V_{ub}|$ from inclusive $B$ decay measurements. Existing estimates would need to be confronted with improved lattice QCD calculations, while the convergence and validity of the OPE in the charm sector needs to be examined carefully. More experimental observables are already available than the total rate and dedicated experimental and theoretical analyses of these are called for.

\section*{Acknowledgement}

The author would like to thank D. Be\'cirevi\'c and P. Gambino for their insightful comments on the manuscript as well as the organizers of the Charm 2009 workshop for the invitation and warm hospitality at this very interesting meeting.

\end{document}